\documentclass[final, aps, pra, floatfix, reprint, nofootinbib, citeautoscript, superscriptaddress, balancelastpage, longbibliography, showkeys]{revtex4-2}
\usepackage[T1]{fontenc}
\usepackage[english]{babel}
\usepackage{amsmath,amssymb,amsthm}
\usepackage{bm}
\usepackage{wasysym}
\usepackage{textcomp}
\usepackage[normalem]{ulem}
\usepackage{tabularx}
\usepackage[colorlinks, urlcolor=blue, citecolor=blue, linkcolor=blue, pdfstartview=FitH]{hyperref}
\usepackage[all]{hypcap}
\usepackage[dvipsnames]{xcolor}
\usepackage{subfigure}
\usepackage{graphicx}
\definecolor{MyColor}{rgb}{0.0, 0.0, 0.0}

\begin{document}

\title{
Collective amplification and anisotropic narrowing of alignment signals in cesium vapor under strong spin exchange near zero magnetic field}

\author{M. V. Petrenko}
\affiliation{Ioffe Institute, 194021 St.~Petersburg, Russia}

\author{A. K. Vershovskii}
\affiliation{Ioffe Institute, 194021 St.~Petersburg, Russia}

\date{\today}

\begin{abstract}
We present the results of an experimental study of the anomalous anisotropy of alignment signals in cesium vapors under strong spin-exchange conditions near zero magnetic field with linearly polarized optical pumping. We show that the anisotropy of the Hanle resonances in the plane perpendicular to the pump beam increases with Cs vapor number density : in one direction the widths remain broadened by spin-exchange, whereas in the other they approach the spin-exchange relaxation free limit. With a further increase in number density , additional nonlinear effects arise, such as signal amplification, bistability, hysteresis, and memory. To explain these effects we construct an illustrative theoretical model incorporating  spontaneous polarization effects under strong spin exchange conditions. The model qualitatively shows that the ultra-narrow alignment resonances may originate from quadrupole anisotropy arising from the projection of spontaneous transverse orientation onto the detection axis. The unique properties of these resonances, such as their extremely small width and magnetic field-controlled bistability with a long-term memory effect, make them promising for use in quantum sensing and information.
\end{abstract}
\maketitle

\section{Introduction}\label{sec:1}

Optically pumped alkali-metal vapors  constitute a versatile platform for quantum sensors that use optically detected magnetic resonance (ODMR), including atomic clocks, magnetometers, and rotation sensors \cite{Budker_Romalis_2007, Kitching_2018, Knappe_Schwindt_Shah_Hollberg_Kitching_Liew_Moreland_2005, Petrenko_Pazgalev_Vershovskii_2021, Petrenko_Pazgalev_Vershovskii_2023, Meyer_Larsen_2014, Vershovskii_Litmanovich_Pazgalev_Peshekhonov_2018}. Their widespread use is due to a combination of unique properties: relatively simple level structure, compatibility with diode-laser pumping across optical and radiofrequency domains, and controllable vapor density via temperature. Efficient optical pumping into nonequilibrium states can be achieved using buffer gases \cite{Franzen_1959} or anti-relaxation coatings \cite{Bouchiat_Brossel_1966}, allowing long spin coherence times.

Most existing optical quantum sensors rely on optical orientation, i.e., the first-order (dipole) moment of the atomic angular momentum, typically generated by circularly polarized light. A major advance in this context is the spin-exchange-relaxation-free (SERF) regime \cite{Happer_Tam_1977, Appelt_Ben-AmarBaranga_Young_Happer_1999}, in which rapid spin-exchange (SE) collisions preserve the total angular momentum and suppress SE broadening. This mechanism enables ultrahigh sensitivity in near-zero magnetic fields \cite{Kominis_Kornack_Allred_Romalis_2003, Ledbetter_Savukov_Acosta_Budker_Romalis_2008}, and has been extended to finite-field operation with partial suppression of SE relaxation \cite{Petrenko_Pazgalev_Vershovskii_2021, Scholtes_Schultze_IJsselsteijn_Woetzel_Meyer_2011, Schultze_Schillig_IJsselsteijn_Scholtes_Woetzel_Stolz_2017}.

Much less attention has been paid to alignment, i.e., the second-order (quadrupole) moment   characterized by zero mean angular momentum. Alignment is produced by linearly polarized light and has been shown to offer several advantages, including the absence of dead zones \cite{Ben-Kish_Romalis_2010, Wang_Wu_Xiao_Wang_Peng_Guo_2021}, reduced heading errors \cite{Hovde_Patton_Versolato_Corsini_Rochester_Budker_2011, Zhang_Kanta_Wickenbrock_Guo_Budker_2023}, and improved long-term stability \cite{Rosner_Beck_Fierlinger_Filter_Klau_Kuchler_Rosner_Sturm_Wurm_Sun_2022}. Theoretical descriptions of alignment dynamics and higher-order polarization moments can be found in \cite{Blum_2012, Omont_1977, Weis_Bison_Pazgalev_2006, Akbar_Kozbial_Elson_Meraki_Kolodynski_Jensen_2024}, with further developments for quasi-stationary regimes reported in \cite{Meraki_Elson_Ho_Akbar_Kozbial_Kolodynski_Jensen_2023}. Alignment-based magnetometry has also been studied outside the SERF regime \cite{Breschi_Weis_2012, LeGal_Lieb_Beato_Jager_Gilles_Palacios-Laloy_2019, Meraki_Elson_Ho_Akbar_Kozbial_Kolodynski_Jensen_2023}.

In recent experiments conducted near zero magnetic field at high temperatures corresponding to high number densities and fast spin exchange, we observed that the alignment resonances exhibit anomalous narrowing \cite{Petrenko_Vershovskii_2025} and a bistable response with hysteresis under elliptical pumping, with memory times reaching hundreds of seconds  \cite{Petrenko_Vershovskii_2026}. These effects indicate that in this system, the quadrupole component of angular momentum is at least partially conserved, despite the fact that it is not protected by the conservation law.

The effects studied in \cite{Petrenko_Vershovskii_2025, Petrenko_Vershovskii_2026} have not yet received a theoretical explanation. To explain the preservation of alignment under strong SE, we proposed the concept of local (on submillimeter scales) alignment-to-orientation conversion (AOC) \cite{Rochester_Ledbetter_Zigdon_Wilson-Gordon_Budker_2012}. As a result, the medium self-organizes into domains with opposite orientations while preserving zero total angular momentum. High buffer gas pressure (several hundred torr of nitrogen) reduces the diffusion rate of cesium atoms, making the existence of such domains possible under continuous pumping. Bistability and hysteresis effects in the presence of a circular component in the pump at the level of tenths of a percent of the total intensity were studied in detail and shown to be explained by the effective field created by oriented atoms. It was shown that in the strong SE regime, the regions of alignment and orientation are capable of coexisting in one cell, and perhaps they are spatially separated.

In this paper, we continue our investigation of these effects. We examine the dependence of the absorption signals and polarization rotation of a linearly polarized beam near zero magnetic field (Hanle resonances in alignment \cite{Breschi_Weis_2012}) on the fields applied in a plane perpendicular to the pump beam. Unlike \cite{Petrenko_Vershovskii_2026}, we do not introduce ellipticity into the pump – all effects are observed with purely linear pumping. First, we show that at relatively low temperatures (on the order of $80^\circ$C), the signal shape corresponds to the theory of \cite{Meraki_Elson_Ho_Akbar_Kozbial_Kolodynski_Jensen_2023}, but with increasing temperature, anisotropy arises and gradually increases in the system. Analysis of the obtained dependences shows that the width of resonances measured while scanning the field perpendicular to the light polarization is determined, as expected, by the SE relaxation rate. However, the width of the resonances obtained by scanning the field along the beam polarization direction corresponds to the widths characteristic of orientation signals in the SERF mode – a phenomenon first observed in \cite{Petrenko_Vershovskii_2025} but not studied in detail. Starting from a certain Cs number density  value corresponding to almost complete absorption of radiation in the cell, new effects arise in the system – namely, bistability and hysteresis. Analysis of the obtained dependences suggests that these effects are caused by a certain internal effective field directed perpendicular to both the beam and the plane of its polarization. The absolute value of this field at a given temperature is independent of external fields, and the sign of its projection onto the same axis is determined by the initial sign of the corresponding component of the external field. The effective field cannot be interpreted as a conventional magnetic field, since it is added to the parallel component of the external field in a nonlinear manner.

To account for the above effects, we introduce a simplified model that takes into account nonlinear spontaneous polarization (SP) described by Fortson and Heckel in \cite{Fortson_Heckel_1987}, which under strong SE can emerge from an arbitrarily small symmetry-breaking seed. 
The main simplification of the model is the reduction of the alignment tensor dynamics to its $\pm1$ components. We show, however, that the model reproduces both the symmetry of alignment signals in the linear regime as well as the experimentally observed nonlinear effects, including anisotropic slowing down of alignment relaxation, bistability, and hysteresis. 

\begin{figure*}[!t]  
\includegraphics[width=\linewidth]{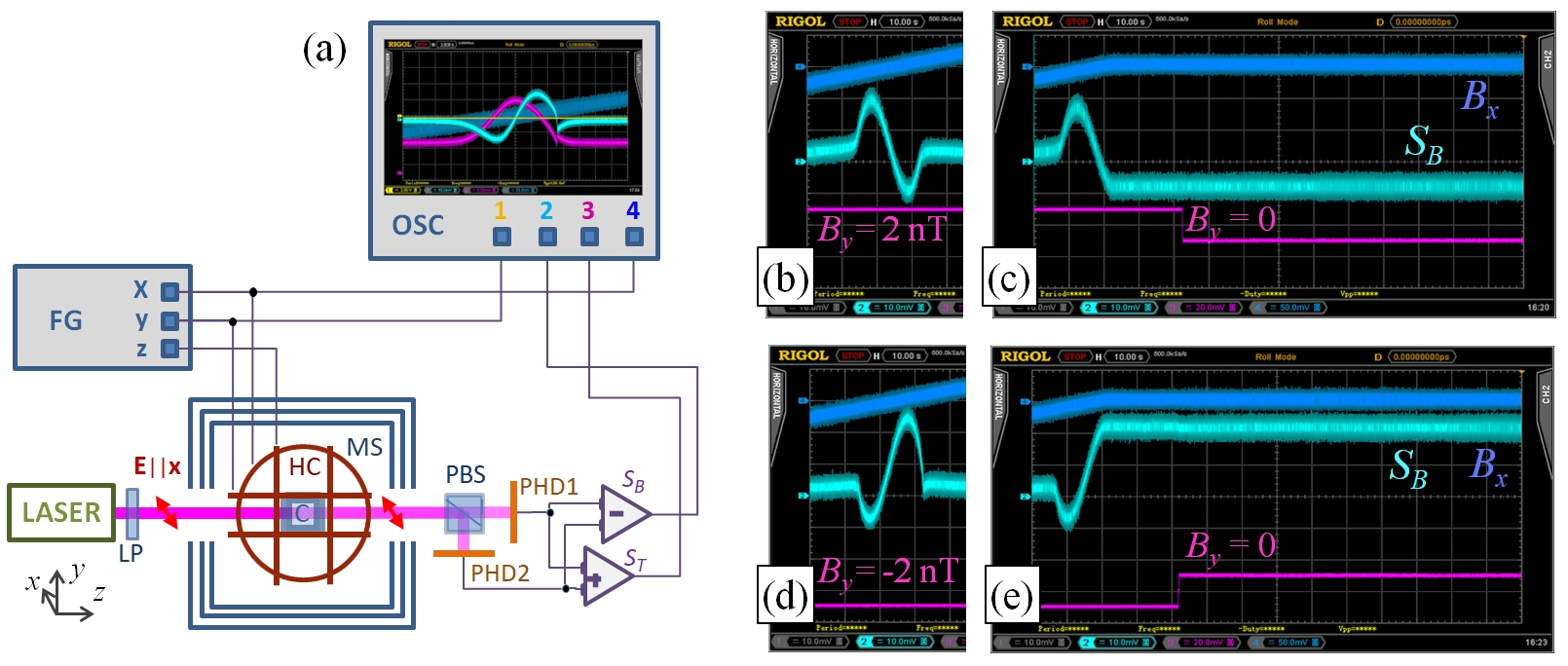}
	\caption{(a) The experimental setup: MS -- magnetic shield, LP -- linear polarizer \textcolor{MyColor}{consisting of a Glan prism and a compensating quarter-wave plate}, HC -- Helmholtz coils system, C -- cell, PBS -- polarization cube, PHD1, PHD2 -- photodiodes, \textcolor{MyColor}{FG -- function generator, OSC -- oscilloscope (input signal assignment is task-dependent). (b)-(e) Oscillograms of signal $S_B$.}}\label{fig1}
\end{figure*}

We emphasize that this model does not claim to be a complete description of the system. Furthermore, the underlying physical picture may not be the only correct one. Its purpose is to demonstrate that the observed phenomena can arise within the standard framework of spin dynamics under strong SE, without invoking additional conservation laws.

The remainder of the paper is organized as follows: Section \ref{sec:2} reviews the basic principles, Section \ref{sec:3} describes the experimental part, Section \ref{sec:4} presents the experimental results and  discusses them. Section \ref{sec:5} presents the theoretical model, and Section \ref{sec:6} discusses the implications and limitations of the proposed model.

\section{Basic principles}\label{sec:2}

We consider optically pumped cesium vapor on the $D_1$ line in the presence of buffer gas nitrogen, which at high enough pressure ($\gtrsim 200$ Torr) ensures efficient quenching of fluorescence, mixing of excited-state sublevels, and pressure broadening of optical transitions. Under these conditions, the hyperfine structure of the excited state is unresolved, and optical pumping occurs simultaneously on multiple transitions.  The atoms are optically pumped by a resonant laser beam propagating along $z$ and linearly polarized along $x$ ($\textbf{E}||\textbf{x}$). Note that this coordinate system differs from that adopted in \cite{Meraki_Elson_Ho_Akbar_Kozbial_Kolodynski_Jensen_2023}: this is due to the fact that in \cite{Petrenko_Vershovskii_2025, Petrenko_Vershovskii_2026} we used both linear and circular polarization, and, accordingly, adopted $\textbf{k}||\textbf{z}$ (Fig.~\ref{fig1}). While circularly polarized light would generate orientation, linearly polarized light produces alignment. In the SERF regime, the collective dynamics of the atomic ensemble can suppress SE relaxation of $orientation$ due to conservation of total angular momentum \cite{Happer_Tam_1977, Appelt_Ben-AmarBaranga_Young_Happer_1999, Kominis_Kornack_Allred_Romalis_2003}.

A quantitative description of alignment can be formulated in the irreducible tensor representation \cite{Blum_2012, Omont_1977}, where the density matrix is expanded in multipoles $\rho_q^{(\kappa)}$. For linearly polarized pumping, the dominant contribution is $\kappa=2$. The stationary solution for the alignment multipoles in a magnetic field, derived in \cite{Meraki_Elson_Ho_Akbar_Kozbial_Kolodynski_Jensen_2023}, leads to explicit expressions for experimentally observable signals.

In the framework of the standard alignment model, the steady-state response is  determined by the dimensionless field components $\omega_i/\Gamma$, where $\Gamma$ is the relaxation rate, and $\omega_i = \gamma B_i$.  After exchanging $x$ and $z$ relative to the notation of \cite{Meraki_Elson_Ho_Akbar_Kozbial_Kolodynski_Jensen_2023} the polarization rotation and transmission signals in linearly polarized light can be written in the form
\begin{equation}
S_B \propto \frac{\Gamma\omega_{z} \left(\Gamma^{2}+4\omega_{x}^{2} +\omega_{yz}^{2} \right)-\omega_{x} \omega_{y} \left(\Gamma^{2}+4\omega_{x}^{2} -2\omega_{yz}^{2}\right)}{\left(\Gamma^{2}+\omega_{x}^{2} +\omega_{yz}^{2} \right)\left(\Gamma^{2}+4\omega_{x}^{2}+4\omega_{yz}^{2}\right)},
\label{eq:m1}
\end{equation}
\begin{equation}
S_T \propto \frac{\Gamma^{2}+ \frac{\left(\omega_{yz}^{2}-2\omega_{x}^{2} \right)^{2}}{\Gamma^{2}} + 2\omega_{yz}^{2} + 5\omega_{x}^{2}}{\left(\Gamma^{2}+ \omega_{x}^{2} + \omega_{yz}^{2} \right)\left(\Gamma^{2}+ 4\omega_{x}^{2} + 4\omega_{yz}^{2}\right)},
\label{eq:m2}
\end{equation}
where $\omega_{yz}^{2}=\omega_{y}^{2}+\omega_{z}^{2}$.
These expressions were obtained in \cite{Meraki_Elson_Ho_Akbar_Kozbial_Kolodynski_Jensen_2023}, and a concise derivation was reproduced in \cite{Petrenko_Vershovskii_2026}. Note that in \cite{Meraki_Elson_Ho_Akbar_Kozbial_Kolodynski_Jensen_2023} a single relaxation rate $\Gamma$ is adopted. When constructing a theoretical model, we will distinguish between the relaxation rate of orientation $\Gamma^{(1)}$ and the relaxation rate of alignment $\Gamma^{(2)}$.

In this work, Eqs.~\eqref{eq:m1}, \eqref{eq:m2} serve as a reference model for the classical alignment response. We will use them to analyze the experimental data and to identify deviations arising in the regime of strong spin exchange, where additional nonlinear effects become important.

\section{Experiment}\label{sec:3}

The experimental setup was essentially identical to that described in \cite{Petrenko_Vershovskii_2025, Petrenko_Vershovskii_2026}. A cell containing several mg of Cs and 200~Torr of $N_2$ was placed in a multilayer magnetic shield (Fig.~\ref{fig1}). The residual field in the shield was compensated by a system of 3D Helmholtz coils. The beam from a VitaWave ECDL laser, tuned to the vicinity of the $F=3 \leftrightarrow F'=4$ transition of the $D_1$ line of Cs, passed through the cell and was detected by a balanced photodetector (Fig.~\ref{fig1}). The pump light power at the cell input was 3~mW with a beam cross section of 0.1~cm$^2$.
\textcolor{MyColor}{Unlike previous studies, the pump-beam ellipticity was maintained at zero with an accuracy better than $0.2^\circ$. To achieve this, a Glan prism was used to provide stable linear polarization, while a compensating quarter-wave plate (omitted from Figure~\ref{fig1} for clarity) compensated small circular polarization components introduced mainly by the cell windows. The presence of such components was monitored by the presence of characteristic signal jumps described in \cite{Petrenko_Vershovskii_2026} when scanning the magnetic field along the $x$-axis. The balanced polarimeter, consisting of a polarizing beam-splitting cube and two photodiodes, was adjusted to provide zero differential DC signal at the operating cell temperature.}

\textcolor{MyColor}{The signals from the photodetector were recorded using a digital oscilloscope.}
To avoid distortion of the recorded signals, field modulation and lock-in detection were not used, resulting in a significantly lower signal-to-noise ratio compared to \cite{Petrenko_Vershovskii_2025, Petrenko_Vershovskii_2026}. 
\textcolor{MyColor}{With photocurrent-to-voltage signal conversion factors in the $S_T$ and $S_B$ channels of 1000 and 5100~V/A, respectively, the signal amplitude was tens of millivolts, resulting in interference from the line frequency and its harmonics. This interference was filtered during primary signal processing using a first-order low-pass digital filter with a time constant of 0.1~s, corresponding to approximately 0.2~nT at the typical scan rate of 2~nT/s. The typical duration of a single recording was 28~s. A linearly varying voltage applied to the magnetic field coil through a resistor was recorded simultaneously with the signal.} 

\textcolor{MyColor}{Figure~\ref{fig1}(d)-(e) shows examples of $S_B$ signal oscillograms, demonstrating both the noise level and the memory effect. At the beginning of the $B_x$ field scan, a field $B_y$ = 2~nT (b), (c) or -2~nT (d, e) was applied to the cell. After recording the complete resonance (b), (d), the scan was repeated; during the second scan (c), (e) the magnetic-field $B_x$ sweep was stopped at the resonance wing and the $B_y$ field was switched off. As a result, the system in both oscillograms appeared to be under the same external influences, but in different internal states. The total recording length is 140 s.}

\begin{figure*}[!t]  
\includegraphics[width=\linewidth]{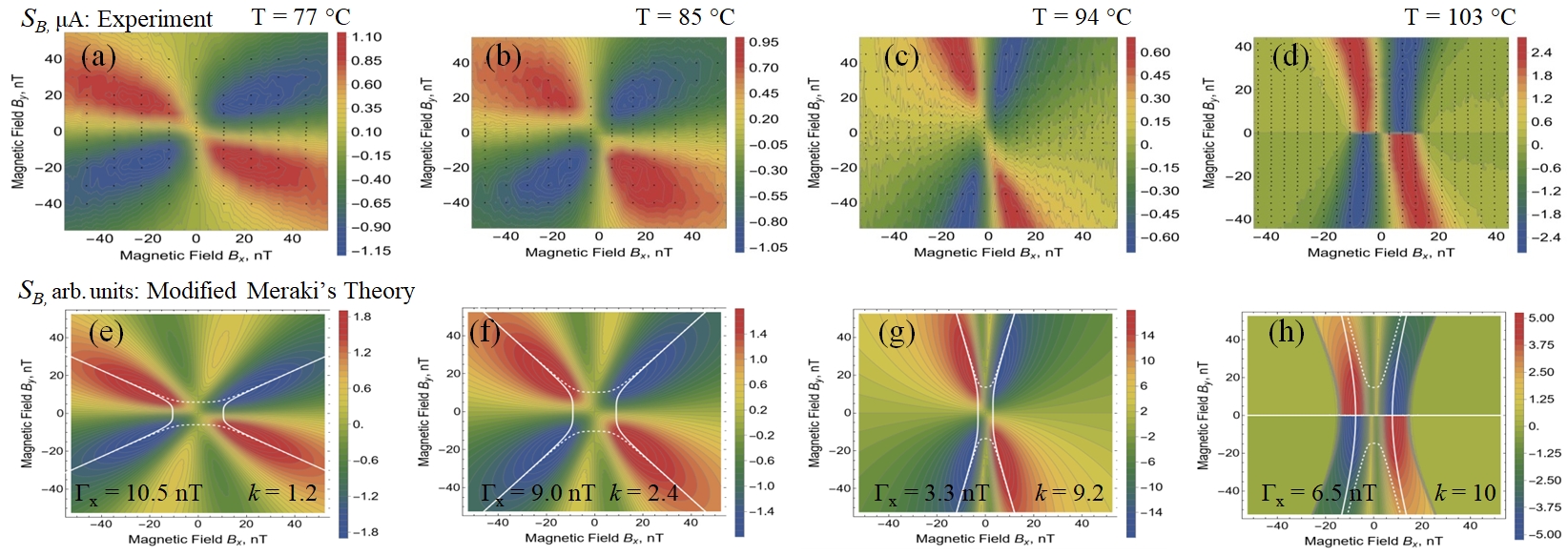}
	\caption{Signals $S_B$ during magnetic field scanning depending on $B_x$, $B_y$ at four temperatures. (a)--(d) Experiment, scanning direction is along $x$. The signal values are expressed in  photocurrent units ($\mu$A). (e)--(h) Modified theory Eq.~\eqref{eq:m1}. Modifications applied to Eqs.~\eqref{eq:m1} of theoretical expressions are described in the text, $k=\Gamma_y/\Gamma_x$. White lines in fragments (e)--(h) indicate approximated dependences of the extrema positions.}
    \label{fig2}
\end{figure*}
\begin{figure*}[!t]  
\includegraphics[width=\linewidth]{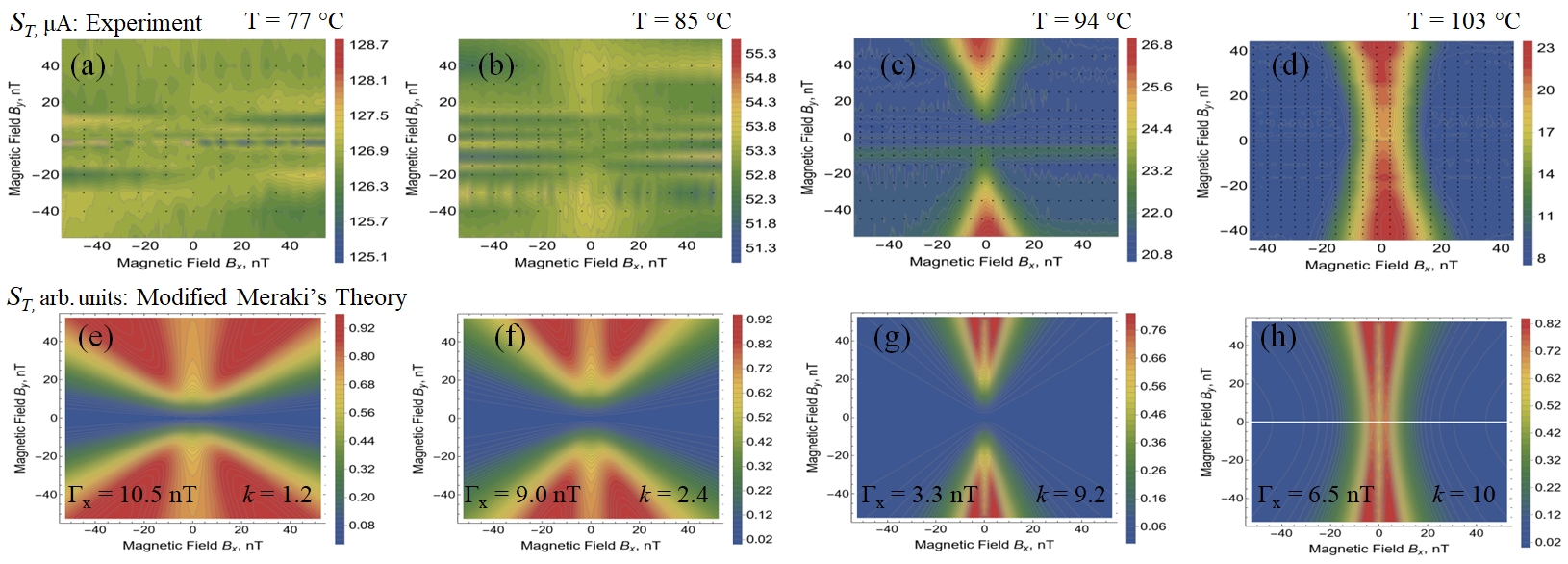}
	\caption{Signals $S_T$ during magnetic field scanning depending on $B_x$, $B_y$ at four temperatures. (a)--(d) Experiment, scanning direction is along $x$. The signal values are expressed in  photocurrent units ($\mu$A). (e)--(h) Modified theory Eq.~\eqref{eq:m2}. Modifications applied to Eqs.~\eqref{eq:m2} of theoretical expressions are described in the text. The signal in fragment (a) is indistinguishable among the noise.}
    \label{fig3}
\end{figure*}

\begin{figure}[!t]  
\includegraphics[width=\linewidth]{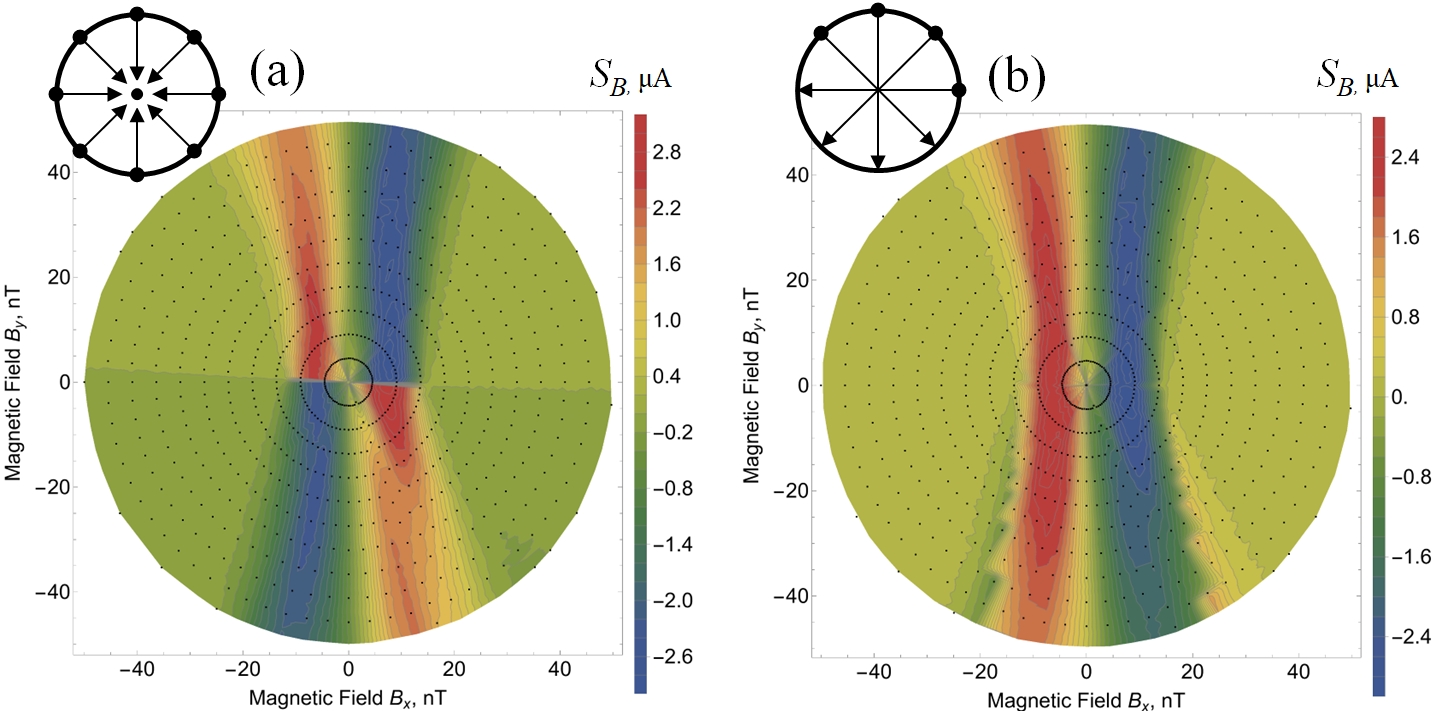}
	\caption{\textcolor{MyColor}{$S_B$} signals. \textcolor{MyColor}{The arrows in the diagrams indicate the scan ranges depicted (see the text for more detailed explanations):} (a) Results of scanning from the edge of a circle with a radius of 42.5~nT to the center. (b) Results of scanning from the upper semicircle to the lower one. The interval between recordings was $5^\circ$. \textcolor{MyColor}{Parts of the same data set} were used in fragments (a) and (b).}
    \label{fig4}
\end{figure}

\begin{figure*}[!t]  
\includegraphics[width=\linewidth]{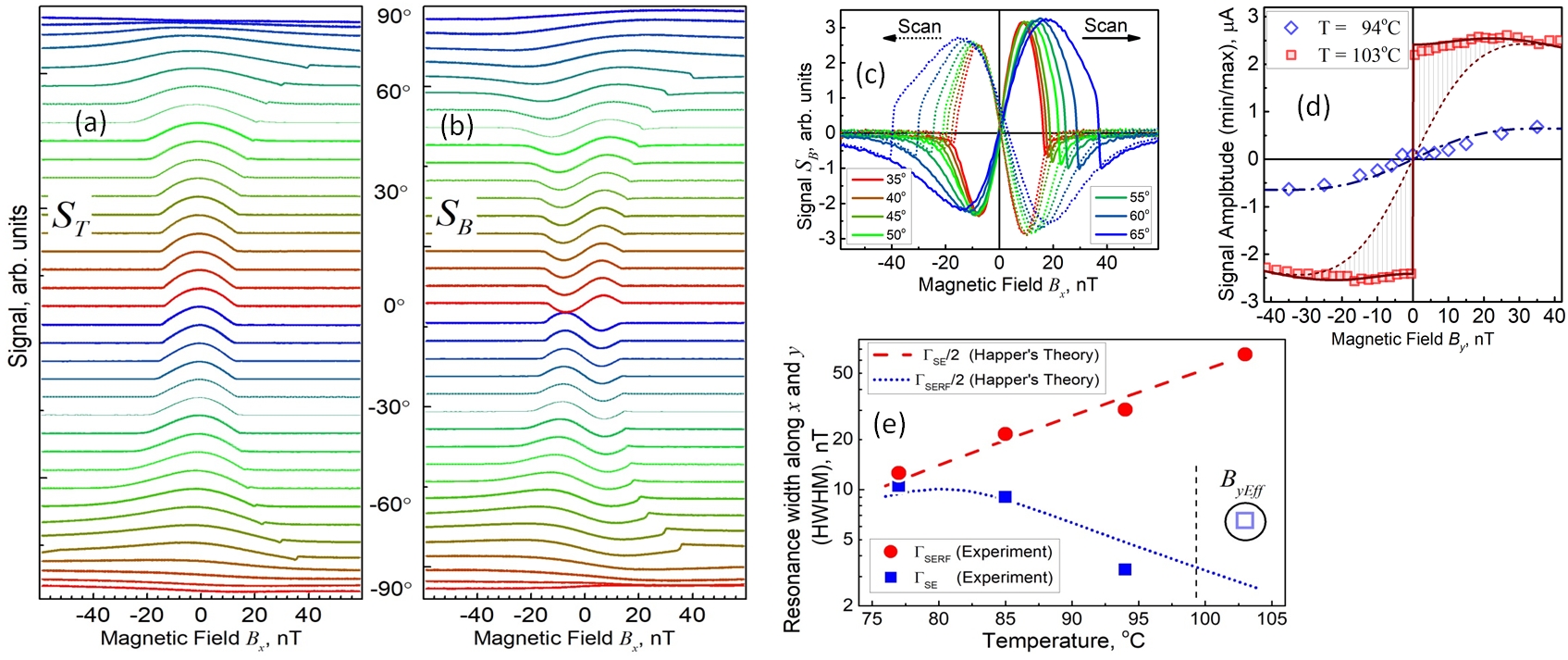}
	\caption{(a), (b), (c) \textcolor{MyColor}{Two-dimensional plots constructed from recordings of the same experimental series as the maps shown in Fig.~\ref{fig4}.} (a) $S_T$. (b), (c) $S_B$. Solid lines present scanning from negative $B_x$ to positive, dots -- vice versa. (d) Dependences of signal amplitudes on the $B_y$ field at two temperatures -- below and above the threshold. Dotted and dash-dotted lines are dispersion curves, solid line is modification of the dispersion curve according to Eq.~\eqref{eq:eff} at $B_{y0}= 23$~nT. (e) Dots -- dependences of the widths (HWHM) of resonances along the $x$ and $y$ axes in Fig.~\ref{fig2}(a)-(d) on temperature; lines -- theoretical width dependences for orientation resonances according to \cite{Happer_Tam_1977} for field $B \gg \Gamma^{(2)}$ (dash) and for field $B=50$~nT ($B$ value in the last case is obtained by fitting). The theoretical HWHM values are divided by two since the widths of the alignment resonances are approximately one half of those of the orientation resonances (see Eqs.~\eqref{eq:m1}, \eqref{eq:m2}).}
    \label{fig5}
\end{figure*}

\textcolor{MyColor}{When scanning along the $x$ axis to construct rectangular maps, the spacing between successive $B_y$ values varied from 2~nT near zero field to 10~nT at larger fields. For circular maps, the scan direction was incremented in $5^\circ$ steps.
When compiling 2D signal alignment maps from continuous recordings, discrete, equally spaced points were extracted and used, shown in black on the maps.}

Since, as follows from Eqs.~\eqref{eq:m1}, \eqref{eq:m2}, the measured resonance width when scanning one field component depends not only on $\Gamma$, but also on the magnetic field components, we, as described above, performed a complete scan of the $B_x$ and $B_y$ fields.
Magnetic-field scanning in the $xy$ plane was performed in the following sequence: first, residual fields were manually compensated based on the system's response to symmetrical changes in the coil field. A fixed value of $B_y$ was then set, and the $B_x$ field was slowly (at a characteristic rate of several nT/s) scanned from negative to  positive values. The procedure was repeated for different values of $B_y$, after which digital filtering of the signal was performed, and a two-dimensional map was constructed. Scanning in the reverse and perpendicular directions was also used to study bistability and hysteresis phenomena. Radial maps were also captured, in which the field was scanned in the radial direction at different angles. These maps were taken at four temperatures ranging from 77 to 103~$^\circ$C.

\section{Experimental Results}\label{sec:4}

The experimental results are presented in Figures~\ref{fig2}--\ref{fig5}. In processing these, we relied on the theory of \cite{Meraki_Elson_Ho_Akbar_Kozbial_Kolodynski_Jensen_2023}, which fully and reliably describes the alignment signals in the classical case. The upper rows of Figs.~\ref{fig2},~\ref{fig3} show the maps obtained by scanning the field along the $x$-axis, as described above. The lower rows show the theoretical maps constructed using Eqs.~\eqref{eq:m1}, \eqref{eq:m2} with modifications described below in this section. The parameters of the theoretical maps were obtained by fitting the corresponding experimental series to the modified Eqs.~\eqref{eq:m1}, \eqref{eq:m2}.

\textcolor{MyColor}{Fig.~\ref{fig4} presents the data obtained at $T=103^\circ$C during radial scanning of the magnetic field in the $xy$-plane over the full angular range of 0-360 degrees relative to the $x$-axis. Since every point of the map is traversed twice during such a scan, the two panels display different subsets of the same experimental data. Fig.~\ref{fig4}(a) contains the first halves of all radial scans, whereas Fig.~\ref{fig4}(b) contains the complete scans recorded for directions between $0^\circ$ and $180^\circ$. In the absence of hysteresis and memory effects the two maps would coincide. Their difference therefore directly reflects the memory effect appearing after the system passes through the zero-field region.}

Fig.~\ref{fig5} shows selected examples of individual records and the results of processing the signals shown in Figs.~\ref{fig2}-\ref{fig4}. Due to the strongly non-analytic shape of the signals Fig.~\ref{fig5}(a),~(b), when processing them, half the distance between the extrema $S_B$ was taken as the resonance width (HWFM).

We compared the resulting map (Fig.~\ref{fig2} -- Fig.~\ref{fig4}) with the maps constructed using Eqs.~\eqref{eq:m1}, \eqref{eq:m2}. The comparison revealed the following patterns: firstly, the symmetry of the obtained signals corresponds to Eqs.~\eqref{eq:m1}, \eqref{eq:m2}, until this symmetry is broken by the effects of bistability and hysteresis.  
Secondly, the shape of the received signals deviates from Eqs.~\eqref{eq:m1}, \eqref{eq:m2} more and more as the temperature and, accordingly, the vapor number density increases.

Up to approximately $100^\circ$C, the observed evolution remains continuous and is manifested primarily by a gradual increase of the anisotropy. Next, a sharp change occurs in the signal behavior: the signal no longer approaches zero at $B_y=0$, instead becoming bistable (as first noted in \cite{ Petrenko_Vershovskii_2026}), and acquiring hysteresis properties (Fig.~\ref{fig4}(a),(b), Fig.~\ref{fig5}(a)--(c)). In addition, the signal decays more rapidly with increasing $|B_{xy}|$. Furthermore, the nonzero-signal region shrinks sharply and acquires very distinct boundaries (Fig.~\ref{fig4}(a),(b), Fig.~\ref{fig5}(a),(b)) -- this is a clear indication of collective nonlinear effects induced by SE.

We modeled these changes by introducing phenomenological corrections into Eqs.~\eqref{eq:m1}, \eqref{eq:m2}: firstly, a scaling factor with respect to $y$, secondly, an internal field $B_{y0}$ such that the effective field perceived by the atoms is equal to 
\begin{equation}
B_{y{\mathrm{Eff}}}=\mathrm{Sign}(B_y)(B_y^2+B_{y0}^2)^{1/2}, 
\label{eq:eff}
\end{equation}
and thirdly, the introduction of a coefficient describing the signal decay at large fields proportional to $|B_{xy}|$, and its zeroing in certain regions. The same phenomenological corrections were applied to the absorption and polarization rotation signals recorded at a given temperature. The expression Eq.~\eqref{eq:eff} is of a purely phenomenological nature. It has an analogue in solid state physics: it corresponds to the phenomenon of \textit{transverse splitting} of EPR/ODMR spectral lines observed in crystals, caused by stress in the crystal lattice. An example of the fit using the signal amplitudes using Eq.~\eqref{eq:eff} for two temperatures (below and above the threshold) is shown in Fig.~\ref{fig5}(d).

 In Fig.~\ref{fig5}(a), (b), the sharp localization of the signal region is clearly visible at small scanning angles relative to $x$ axis, as well as the hysteresis at large scanning angles. Fig.~\ref{fig5} (c) shows the  inversion and hysteresis of the signals taken at the opposite scanning direction; as shown in \cite{ Petrenko_Vershovskii_2026}, the times at which the system can remain trapped in either of the two states reach hundreds of seconds or more, indicating that these states are self-sustaining. Fig.~\ref{fig5} (d) shows the dependences of the signal amplitudes on the $B_y$ field and their approximation without and with the effective field $B_{y0}$. Fig.~\ref{fig5}(e) compares the experimentally measured linewidths of the resonances of Fig.~\ref{fig2} measured along the $x$ and $y$ axes, with the theoretical linewidths of the orientation resonances according to \cite{Happer_Tam_1977} for a strong and weak magnetic field. The theoretical linewidths are divided by two, since the widths of the alignment resonances are approximately half the widths of the orientation resonances Eqs.~\eqref{eq:m1}. From Fig.~\ref{fig5}(e) it can be concluded that the width of the resonances along $y$ corresponds very well to the SE width of the resonance, while the width of the resonances below the threshold temperature approaches the SERF-limited width for the orientation. Resonances above the critical temperature are slightly broadened by the effective field $B_{y0}$.

\textcolor{MyColor}{The relative amplitude of the polarization-rotation signal, which in the first approximation is proportional to the polarization-rotation angle per Cs atom, is calculated as $S_{nB}=S_{mB}/(I_{ph}\cdot n_{Cs})$, where $S_{mB}$ is the maximum amplitude of $S_B$ signal (Fig.~\ref{fig2}), $I_{ph}$ is the photocurrent (see Fig.~\ref{fig3}), and $n_{Cs}$ is the Cs vapor number density.} Up to a threshold of $\approx100^\circ$C ($T=77-94^\circ$C) \textcolor{MyColor}{$S_{nB}$} remains almost unchanged with temperature and lies within the range of $S_B/I_{ph}/n=2.9\pm 0.4 \times 10^{-15}$~cm$^3$. However, after exceeding the threshold, at $T=103^\circ$C, it increases  to $7.1 \times 10^{-15}$~cm$^3$. The peak-to-peak amplitude of the $S_T$ signal at this temperature is comparable to the total photocurrent, and the amplitude of the $S_B$ signal is about 1/3 of it.

\section{Theoretical Model}\label{sec:5} 

We consider a system described in Section \ref{sec:2}. We assume that the temperature is sufficiently high that SE collisions dominate, and we operate in the SERF regime where the Larmor frequency $\omega_L = \gamma |\mathbf{B}|$ is much smaller than the SE rate $R_{SE}$ ($\gamma$ is the gyromagnetic ratio, for Cs $\gamma = 3.5$~Hz/nT).

We expand the ground-state density matrix in irreducible tensor operators $\rho_q^{(\kappa)}$ (\cite{Omont_1977,Happer_Tam_1977}). The relevant ranks are $\kappa=1$ (orientation, dipole) and $\kappa=2$ (alignment, quadrupole).  
\textcolor{MyColor}{We choose the quantization axis along the $x$ axis ($\textbf{E}||\textbf{x}$). Alignment may then be represented either by the
spherical tensor components $\rho_q^{(2)}$ or by a real symmetric
traceless Cartesian tensor $\textbf{Q}$ \cite{Omont_1977}:
\begin{align}
\rho_{0}^{(2)}
&=
\frac{1}{\sqrt6}
\left(
2Q_{xx}-Q_{yy}-Q_{zz}
\right),
\\
\rho_{\pm1}^{(2)}
&=
\mp\frac12
\left[
(Q_{xy}+Q_{yx})
\pm
i(Q_{xz}+Q_{zx})
\right],
\\
\rho_{\pm2}^{(2)}
&=
\frac12
\left[
(Q_{zz}-Q_{yy})
\pm
i(Q_{yz}+Q_{zy})
\right].
\end{align}
where the numerical coefficients are given for one commonly used normalization of the irreducible tensor operators.}

\textcolor{MyColor}{Throughout this section we assume $B_z=0$.
Since the optical pumping establishes a stationary alignment along the
$x$ axis, only three tensor components participate in the present
dynamics: the stationary diagonal component $Q_{xx}$ and the
off-diagonal components $Q_{xy}$ and $Q_{zx}$.}
\textcolor{MyColor}{We therefore introduce the real variables $A_x\propto Q_{xx}$, $A_y\propto Q_{xy}$, and $A_z\propto Q_{zx}$, so that
\begin{align}
\rho_{-1}^{(2)}&=A_z+iA_y,\\
\rho_{+1}^{(2)}&=-A_z+iA_y.
\end{align}
According to the symmetry analysis of Meraki \emph{et al.}~\cite{Meraki_Elson_Ho_Akbar_Kozbial_Kolodynski_Jensen_2023},
\begin{equation}
S_B\propto
i\left(\rho_{+1}^{(2)}+\rho_{-1}^{(2)}\right)\propto A_y,
\end{equation}
In the present simplified model the alignment dynamics can be written as}
\begin{align}
\dot A_y &= \omega_xA_z-\Gamma^{(2)}A_y,
\\
\dot A_z &= \omega_yA_x-\omega_xA_y-\Gamma^{(2)}A_z,
\label{eq:dot_Az}
\end{align}
\textcolor{MyColor}{where the stationary alignment $A_x$ created by optical pumping is
\begin{equation}
A_x=\frac{R}{\Gamma^{(2)}},
\end{equation}
with $R$ being the optical pumping rate and $\Gamma^{(2)}$ the alignment relaxation rate.}

In the steady state,
\begin{equation}
A_z=
\frac{
\Gamma^{(2)}\omega_y A_x
}{
\Gamma^{(2)2}+\omega_x^2
}
\label{eq:As_lin}
\end{equation}
and
\begin{equation}
A_y=
\frac{
\omega_x\omega_y A_x
}{
\Gamma^{(2)2}+\omega_x^2
}.
\label{eq:Ac_lin}
\end{equation}
These expressions describe the conventional linear alignment response with the characteristic width determined by $\Gamma^{(2)}$.

\textcolor{MyColor}{We now extend the linear alignment model by introducing a phenomenological description of spontaneous polarization (SP).} We assume that linearly polarized light, interacting with a medium in a transverse magnetic field, creates not only alignment, but also a small first-order angular momentum that favors the development of SP along the $z$ axis under strong SE conditions. We also assume that this symmetry breaking initially generates the orientation Cartesian component $P_z$, while the component $P_y$ appears through Larmor precession in the field $B_x$.
This assumption is supported by the experimental observations reported in \cite{Petrenko_Vershovskii_2026}: the sign of the alignment signal is primarily determined by the fields $B_x$ and $B_y$, but does not depend on $B_z$. 
The resulting conversion chain is

\begin{equation}
A_x
\xrightarrow{\omega_y}
A_z
\xrightarrow{\mathrm{SP}}
P_z
\xrightarrow{\omega_x}
P_y.
\end{equation}

The orientation dynamics is described phenomenologically by
\begin{align}
\dot P_y &= \omega_x P_z - \Gamma^{(1)}P_y,
\label{eq:Py_full}
\\
\dot P_z &= -\omega_x P_y
-\Gamma^{(1)}P_z
+\chi P_z
-\eta P_z^3
+\alpha A_z .
\label{eq:Pz_full}
\end{align}

Here $\Gamma^{(1)}$ is the orientation relaxation rate, $\chi$ is the spontaneous-polarization gain coefficient, $\eta>0$ describes nonlinear saturation, and $\alpha$ characterizes the efficiency with which the alignment anisotropy seeds SP. Saturation is introduced by a third-order term, which is the first nonlinear term of the Taylor series expansion, consistent with the required symmetry. The angular momentum transferred to $P_z$ arises as a result of optical pumping, rather than from the seed itself, so SP does not require additional  relaxation of $A_z$ (the opposite case will be further discussed at the end of this section).

In the linear stationary regime ($\eta=0$), the steady-state solution of Eqs.~\eqref{eq:Py_full}--\eqref{eq:Pz_full} gives
\begin{equation}
P_y=
\frac{
\omega_x
}{
\Gamma^{(1)}
}
P_z,
\label{eq:PyPz}
\end{equation}
and
\begin{equation}
P_z=
\frac{
\alpha\Gamma^{(1)}A_z
}{
\Gamma^{(1)2}
-\chi\Gamma^{(1)}
+\omega_x^2
}.
\label{eq:Pz_As}
\end{equation}

Substituting Eq.~\eqref{eq:As_lin}, we obtain
\begin{equation}
P_z
=
\frac{
\omega_y\alpha\Gamma^{(1)}\Gamma^{(2)}A_x
}{
\left(
\Gamma^{(2)2}+\omega_x^2
\right)
\left(
\Gamma^{(1)2}
-\chi \Gamma^{(1)}
+\omega_x^2
\right)
},
\label{eq:Pz_linear}
\end{equation}
and
\begin{equation}
P_y
=
\frac{
\omega_x\omega_y\alpha\Gamma^{(2)}A_x
}{
\left(
\Gamma^{(2)2}+\omega_x^2
\right)
\left(
\Gamma^{(1)2}
-\chi \Gamma^{(1)}
+\omega_x^2
\right)
}.
\label{eq:Py_linear}
\end{equation}

$P_y$ has exactly the same symmetry
$P_y\propto\omega_x\omega_y$,
as the alignment signal $A_y$. However, unlike $A_y$, the characteristic width of $P_y$ is determined by the slow orientation relaxation rate $\Gamma^{(1)}$ rather than by the fast alignment relaxation rate $\Gamma^{(2)}$.

\begin{figure}[!t]  
\includegraphics[width=\linewidth]{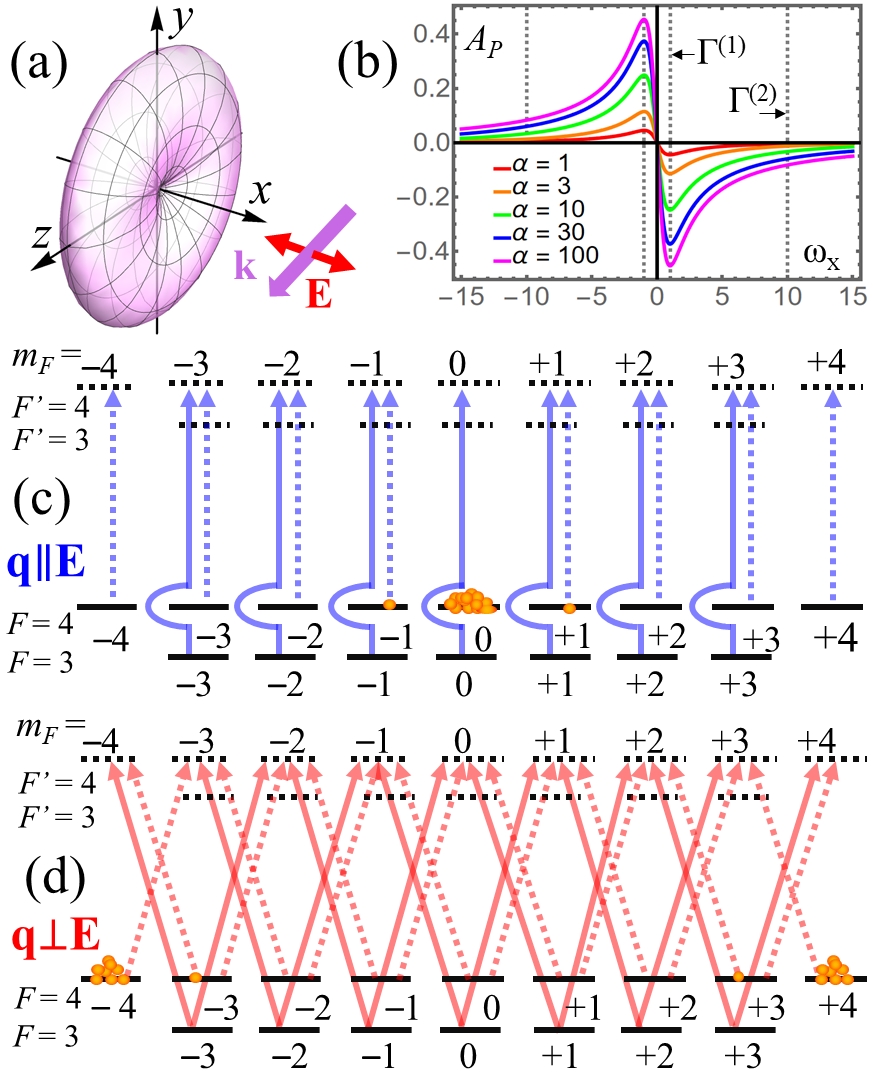}
	\caption{(a) Spatial distribution of angular momentum under pumping with linearly polarized light, corresponding to negative alignment along $x$ and positive alignment along $y$ and $z$ ("donut") \cite{Rochester_Ledbetter_Zigdon_Wilson-Gordon_Budker_2012}. (b) Dependences of the $A_P$ component (arb. units) on the normalized magnetic field $\omega_x = B_x/\gamma$ for different AOC coefficients $\alpha$ according to Eq.~\eqref{eq:Ap_AOC}: $\Gamma^{(1)}=1,\Gamma^{(2)}=10$. (c), (d) Schemes of pumping with linearly polarized light resonant to the $F = 3 \rightarrow F' = 4$ transition: (c) the quantization axis is parallel to the electric vector $\mathbf{E}$ of the light wave, the alignment in $F = 4$ is negative; (d) the quantization axis is perpendicular to $\mathbf{E}$, the alignment is positive. The dotted lines indicate additional pumping due to the overlap of broadened optical spectral profiles.}
\label{fig6}
\end{figure}

We assume that the pumping creates a negative alignment along  $x$ dominated by the sublevel $m_x=0$ (this assumption will be substantiated in Section \ref{sec:6}). This distribution corresponds to positive alignment with respect to the transverse axes $y$ and $z$. The angular momentum distribution has a "donut" shape (Fig.~\ref{fig6}(a)) with the symmetry axis parallel to $\mathbf{E}$ (see, for example, \cite{Rochester_Ledbetter_Zigdon_Wilson-Gordon_Budker_2012}).

Thus, negative alignment created by the pump along $x$ automatically implies an enhanced population of stretched states with large $|m_y|$ and $|m_z|$. These transverse stretched states can act as a basis for SP under strong SE.
Rapid SE collisions may collectively synchronize atomic spins inside local domains, allowing one of the two stretched states ($m_y=+F$ or $m_y=-F$) to become populated depending on the dominant symmetry-breaking factor, uniquely related to the sign of the transverse field $B_y$. Under this assumption, the coefficient $\alpha$ characterizes the susceptibility of collective SP to the transverse quadrupole anisotropy $A_z \propto B_y$ (Eq.~\eqref{eq:As_lin}) already present in the pumped ensemble.

\textcolor{MyColor}{In the limiting case where the alignment is completely determined by the
SP, the proportionality}
\begin{equation}
A_P=\xi P_y
\label{eq:Ap}
\end{equation}
\textcolor{MyColor}{follows directly from the properties of the rank-2 alignment tensor
associated with a fully oriented atomic state} (for details, see Appendix~\ref{appendix:A}).
Using Eq.~\eqref{eq:Py_linear}, we obtain
\begin{equation}
A_P
=
\frac{
\omega_x\omega_y\alpha\xi\Gamma^{(2)}A_x
}{
\left(
\Gamma^{(2)2}+\omega_x^2
\right)
\left(
\Gamma^{(1)2}
-\chi \Gamma^{(1)}
+\omega_x^2
\right)
}.
\label{eq:Ap_final}
\end{equation}

The total experimentally measured signal can therefore be written as
\begin{equation}
S_B \propto A_y+A_P.
\label{eq:S_total}
\end{equation}

Although both contributions have the same sign, corresponding to negative alignment, and have the same symmetry with respect to magnetic-field reversal, their physical origin is fundamentally different. The component $A_y$ is a true rank-2 alignment component relaxing with the rate $\Gamma^{(2)}$, whereas $A_P$ is not an independently evolving tensor quantity. In the limit $\chi \ll \Gamma^{(1)} \ll \Gamma^{(2)}$  its dynamics is fully determined by $\Gamma^{(1)}$.
In the SERF regime the $A_P$ contribution relaxes significantly more slowly than $A_y$ and therefore dominates the signal.

The condition for self-sustaining SP follows from the instability of the linearized orientation dynamics:
\begin{equation}
\chi>\Gamma^{(1)}.
\label{eq:SP_cond}
\end{equation}

The singularity in Eq.~\eqref{eq:Ap_final} at
$\chi=\Gamma^{(1)}$
appears because the present model is constructed within the linear stationary approximation. In particular, the nonlinear saturation term $-\eta P_z^3$ was omitted in deriving Eq.~\eqref{eq:Ap_final}. Near and above the threshold, this term becomes essential and stabilizes the dynamics. In the limit $\omega_x \rightarrow 0$, $\omega_y \rightarrow 0$, Eq.~\eqref{eq:Pz_full} yields
\begin{equation}
\eta P_z^3+\left(\Gamma^{(1)}-\chi
\right)P_z=0.
\label{eq:Landau}
\end{equation}
When condition Eq.~\eqref{eq:SP_cond} is met, two non-trivial stable stationary solutions exist even at vanishing external transverse field:
\begin{equation}
P_z=
\pm
\sqrt{
\frac{
\chi-\Gamma^{(1)}
}{\eta}
}.
\label{eq:Pz_spont}
\end{equation}

The emergence of two equivalent states with opposite signs of $P_z$ is the standard signature of spontaneous symmetry breaking: the system retains one of the two possible directions of collective orientation even in the absence of an external orienting field. Since both states are locally stable, the final state depends on the previous evolution of the system and on weak perturbations acting as a seed. As a result, slow scans of the magnetic field may lead to hysteresis and long-lived memory effects, because transitions between the two branches require overcoming the nonlinear stability barrier imposed by the saturation term.

The threshold condition Eq.~\eqref{eq:SP_cond} also allows us to qualitatively explain the sharp borders of the signal regions (Fig.~\ref{fig2}(d), Fig.~\ref{fig5}(a), (b)): outside these regions the effective gain becomes insufficient for self-sustaining SP.
Note that as the temperature increases, $\Gamma^{(1)}$ decreases, and condition Eq.~\eqref{eq:SP_cond} is more easily fulfilled.

The effect responsible for the appearance of orientation may in principle be not only SP. Other mechanisms, such as unidirectional AOC, may also be incorporated phenomenologically into the model. In the simplest approximation, this can be done by introducing an additional dissipative term $-\alpha A_z$ into Eq.~\eqref{eq:dot_Az}, corresponding to that in Eq.~\eqref{eq:Pz_full}. 
This additional term can describe not only AOC, but also incoherent SP accompanied by dissipation.
In this case, the expression Eq.~\eqref{eq:Ap_final} retains its general structure, while its first denominator changes.
 In the limit $\chi \rightarrow 0$,
 $\eta \rightarrow 0$ it takes the form
\begin{equation}
A_P
=
\frac{
\omega_x\omega_y
\alpha\xi\Gamma^{(2)}A_x
}{
\left[
\Gamma^{(2)}
\left(
\Gamma^{(2)}+\alpha
\right)
+
\omega_x^2
\right]
\left(
\Gamma^{(1)2}
+\omega_x^2
\right)
}.
\label{eq:Ap_AOC}
\end{equation}
As a result, the broad wings of the resonance become additionally broadened and smoothed, whereas the narrow central structure is still governed by the slow orientation relaxation rate $\Gamma^{(1)}$ (Fig.~\ref{fig6}(b)).
It should be noted, however, that AOC alone is not sufficient to create the self-sustaining orientation required for the hysteresis and memory (Fig.~\ref{fig2}(d), Fig.~\ref{fig4}(b)).

\begin{figure*}[!t]  
\includegraphics[width=\linewidth]{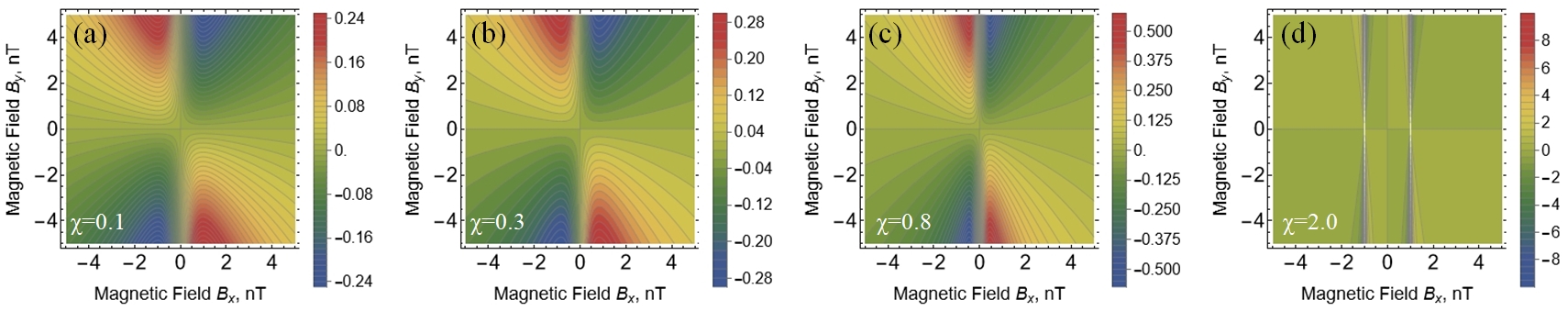}
	\caption{\textcolor{MyColor}{Dependences of the $A_P$ component (arb. units)  on the normalized magnetic fields $\omega_x$, $\omega_y$ according to Eq.~\eqref{eq:Ap_final} at $\Gamma^{(1)}=1$,$\Gamma^{(2)}=10$, $\alpha\xi A_x=1$. (a) $\chi=0.1$, (b) $\chi=0.3$, (c) $\chi=0.8$, (d) $\chi=2.0$.}}
\label{fig7}
\end{figure*}

\textcolor{MyColor}{The numerical solutions also show that AOC alone does not produce the anisotropy observed experimentally. Within the present model, the anisotropy is controlled by the parameter $\chi$, which characterizes the onset of SP. Figure~\ref{fig7} presents the calculated distributions of the alignment component $A_P$ (arb.~units), obtained from Eq.~\eqref{eq:Ap_final}, as functions of the normalized magnetic-field components $\omega_x=B_x/\gamma$ and $\omega_y=B_y/\gamma$ for $\chi$=0.1, 0.3, 0.8, 2.0. The remaining parameters were $\Gamma^{(1)}=1$, $\Gamma^{(2)}=10$, and $\alpha\xi A_x=1$.}

\textcolor{MyColor}{As seen from Fig.~\ref{fig7}, increasing $\chi$ leads not only to progressive resonance narrowing but also to an increase in the signal, the emergence of pronounced anisotropy and of the characteristic four-lobed symmetry. These features are in good qualitative agreement with the experimental maps shown in Fig.~\ref{fig2}(a)-(c). For $\chi\ge1$ (Fig.~\ref{fig7}d), the system leaves the linear regime assumed in the present model. Although the theory is no longer expected to provide a quantitative description in this regime, the calculated distributions still reproduce the principal symmetry features observed experimentally (cf. Fig.~\ref{fig2}(d)).}

\textcolor{MyColor}{These calculations demonstrate that the introduction of spontaneous polarization is sufficient to reproduce the principal qualitative features observed experimentally, including signal amplification, resonance narrowing, anisotropy, the characteristic four-lobed symmetry, and the transition toward the nonlinear regime as the SP gain parameter increases.}

\section{Discussion}\label{sec:6}

Thus, the new experimental data confirm the effects reported previously in \cite{Petrenko_Vershovskii_2025, Petrenko_Vershovskii_2026}. We showed that, with increasing temperature (and therefore atomic number density ), the system progressively departs from the classical regime described in \cite{Meraki_Elson_Ho_Akbar_Kozbial_Kolodynski_Jensen_2023}. At moderate density, this evolution manifests itself primarily through a strong anisotropy of the resonance widths: when the magnetic field is scanned perpendicular to the light polarization vector $\mathbf{E}$, the resonance width remains determined by the fast SE relaxation, whereas for scans along $\mathbf{E}$ the resonance width approaches the narrow SERF limit. 

Upon further increase of temperature and SE rate, qualitatively new nonlinear effects emerge: effective-field-like behavior, hysteresis and bistability. The simplified theoretical model proposed in this paper suggests that the observed narrow resonances may originate from alignment dynamics.
However, this alignment is not an independent dynamical variable, but rather a feature of the transverse orientation, while this orientation itself is seeded by the alignment through the SP effect. 

At relatively low atomic density, SP is entirely determined by the external seed, and the system behaves as a linear amplifier of alignment signals. In this regime, the orientation survives only while sustained by the seed generated through alignment and magnetic-field-induced precession. A SERF-like narrowing of resonances along one axis is observed. A clear distortion is visible in Fig.~\ref{fig2}(a), where the resonance petals become weakly bent. Such bending is absent in the linear model  \cite{Meraki_Elson_Ho_Akbar_Kozbial_Kolodynski_Jensen_2023} (Fig.~\ref{fig2}(e)). It reflects the fact that, at relatively low temperatures, the system leaves the \textcolor{MyColor}{linear} regime already at modest values of $|B_y|$, producing an additional nonlinear broadening along the $x$ direction.

As the temperature and SE rate increase, an increasing fraction of the atomic ensemble enters the \textcolor{MyColor}{nonlinear regime. Above a threshold} the polarization can sustain itself even in the absence of an external seed. Moreover, the established polarization state can persist even after the seed, i.e. alignment, changes sign (Fig.~\ref{fig4}(b)). This explains the emergence of hysteresis and bistability effects.

This nonlinear transition is also reflected in the signal amplitude.
Below the threshold the signal amplitude, normalized to atomic number density  and optical power, remains nearly constant. However, after entering the strongly nonlinear regime, the normalized amplitude increases. Because of the SP effect, the medium acts as an effective amplifier of the alignment response, preserving its symmetry properties while providing anisotropic suppression of relaxation. 

An important open question is whether the experimentally observed signals originate purely from the quadrupole component or may also contain contributions from the orientation component $P_z$. The $P_z$  component can be observed in the polarization rotation signal due to circular birefringence. However, orientation along $z$ cannot directly generate linear dichroism for light propagating along the same axis. In our experiment, narrow structures are observed not only in polarization rotation, but also directly in transmission, and their shape closely follows the behavior expected for alignment signals (Fig.~\ref{fig3}). Moreover, as shown in Section \ref{sec:4}, at high temperature the amplitude of the transmission signal exceeds the amplitude of the polarization rotation signal and is comparable to the total photocurrent. Therefore, we believe that circular birefringence cannot make a dominant contribution to the observed effects.

We suggest that the experimentally observed signal is determined by the superposition
$S_B\propto A_y+A_P$, where $A_y$ is the ordinary rapidly relaxing alignment contribution, and $A_P$ is the projection of transverse orientation $P_y$ onto the detection axis $\mathbf{x}||\mathbf{E}$. 
$A_P$ inherits from $P_y$ the slow relaxation rate of orientation, therefore it dominates in the \textcolor{MyColor}{nonlinear} regime and produces \textcolor{MyColor}{anisotropic} narrow resonances.

The proposed model eliminates the need for a real internal magnetic field generated by neighboring oriented atoms: the orientation no longer affects the alignment through an effective magnetic field; instead, the nonlinear feedback acts directly within the orientation subsystem through SE. 

The model also allows conclusions regarding the dominant optical pumping mechanism to be drawn under conditions of strongly overlapping optical spectral profiles. When the laser is tuned near the $F=3 \rightarrow F'=4$ transition of the Cs $D_1$ line, this transition dominates not only because of spectral proximity, but also because its oscillator strength is approximately three times larger than that of the nearby $F=3 \rightarrow F'=3$ transition. Under linearly polarized excitation, atoms are efficiently pumped from the $F=3$ manifold into the $F=4$ manifold, forming negative alignment along $\mathbf{E}$ and positive alignment in the transverse $yz$ plane (Fig.~\ref{fig6}(c),~(d)). The subsequent spontaneous orientation therefore develops directly within the $F=4$ manifold, where the SERF dynamics is strongest. Additional optical pumping in $F=4$ arises due to the overlap of broadened optical spectral profiles and due to rapid SE transfer between the two hyperfine manifolds. 

Detection of linear dichroism signals is likewise enabled by overlap of the broadened optical profiles, but may additionally involve partial transfer of alignment back to the $F=3$ manifold via rapid spin-exchange collisions. This latter process could in principle be the cause of positive feedback, but we do not have sufficient data to examine this possibility.

A related unresolved issue concerns the mechanism by which an extremely weak transverse field $B_y$ \textcolor{MyColor}{(of the order of nanotesla, as illustrated by the signal behavior in  Fig.~\ref{fig1}(c)-(e))} selects the direction of SP. Similar behavior was previously observed in \cite{Petrenko_Vershovskii_2026}, where the sign of the polarization was determined by a deliberately introduced weak (tenths of a percent) ellipticity of the pumping light, in agreement with \cite{Fortson_Heckel_1987}. In the present experiment the pump light is linearly polarized. However, rotation of the alignment tensor by the magnetic field in the presence of nonlinear effects may lead to unequal absorption of the two circular components of the linearly polarized pump. Such a mechanism would effectively generate a weak circular seed and reduce the problem to the situation experimentally realized in \cite{Petrenko_Vershovskii_2026}.
\textcolor{MyColor}{Another possibility is that in the absence of circularly polarized pumping, SP may adiabatically follow the magnetic field, similarly to the behavior discussed in \cite{Klipstein_Lamoreaux_Fortson_1996}. In this case, the experimentally observed symmetry of the $S_B$ signal is also preserved (see Appendix~\ref{appendix:A}).}

\textcolor{MyColor}{Above we did not consider either the transverse intensity distribution of the pump beam or its spatial evolution inside the cell. In an experiment, the conditions for the onset of SP may be satisfied only within a limited region of the cell. This explains why the magnitude of the SP-related effects can vary smoothly with changes in the experimental parameters, rather than reaching their maximum values.}

\textcolor{MyColor}{Maintaining SP under linearly polarized optical pumping requires a mechanism for removing the excess angular momentum transferred in the opposite direction. Molecular nitrogen appears to be the most likely candidate for the role of mediator in this process, which may explain its previously noted special role in SP phenomena \cite{Andalkar_Warrington_Romalis_Lamoreaux_Heckel_Fortson_2002}.}

Alternative interpretations of the observed phenomena may involve mechanisms such as bidirectional AOC, self-organization of oppositely oriented domains, or conservation of quadrupole angular-momentum moments under rapid spin-exchange.  However, at present we do not find sufficient grounds to conclude that any of these mechanisms provides a more natural or self-consistent explanation of the experimental observations than the model proposed here. In particular, none of these alternative scenarios straightforwardly explains the simultaneous emergence of all the effects observed in the strongly nonlinear regime.

\section{Conclusions}
We have experimentally investigated alignment signals in Cs vapor under conditions of strong spin exchange and linearly polarized optical pumping near zero magnetic field. The measurements reveal an explicit anisotropy of the magnetic response: resonances obtained by scanning the field along the pump-polarization axis exhibit strong narrowing characteristic of the SERF regime. \textcolor{MyColor}{Up to approximately $100^\circ$C, the observed evolution remains continuous and is manifested primarily by a gradual increase of the anisotropy.}
At higher temperature, the system additionally demonstrates effective-field-like behavior, bistability, hysteresis, and signal amplification.

To interpret these observations, we developed a minimal model combining spontaneous polarization, nonlinear saturation, and the tensor structure of the atomic angular-momentum distribution under rapid SE mixing.
We suggest that the experimentally observed narrow resonances arise from quadrupole anisotropy, i.e. alignment, associated with transverse orientation projected onto the detection axis. Although these resonances retain symmetry and alignment signal detection properties, their slow dynamics are determined by the orientation subsystem.

\textcolor{MyColor}{The theoretical model reproduces the principal qualitative features observed experimentally, including signal amplification, resonance narrowing, anisotropy, and characteristic four-lobed symmetry. The predicted structure is simpler than that of the conventional alignment theory while remaining in very good qualitative agreement with the experimental observations. The model also captures the transition toward the nonlinear regime above the spontaneous polarization threshold.}

At the same time, the mechanism by which extremely small transverse magnetic fields select the direction of spontaneous polarization remains an open question requiring further investigation.

We believe that our results provide insight into nonlinear collective effects in dense alkali metal vapors. Moreover, the unique properties of SERF-like alignment resonances, such as their extremely small width and magnetic-field-controlled bistability with memory times reaching hundreds of seconds, make them promising for applications in quantum sensing and information technologies.

\section*{Acknowledgments}

The authors thank Prof. E. Aleksandrov and Dr. A. Pazgalev for valuable discussions. This research was funded by the baseline project FFUG-2024-0039 at the Ioffe Institute.

\appendix

\section{\textcolor{MyColor}{Alignment tensor associated with a fully oriented state}}
\label{appendix:A}
\addcontentsline{toc}{section}{Appendix}

Consider atoms polarized along the $y$ axis in the SERF regime, so that the ensemble approaches the stretched state. Expanding this state in the basis $|F,m_x\rangle$ yields
\begin{equation}
|F,F\rangle_y
=
\sum_{m_x=-F}^{F} c_{m_x}|F,m_x\rangle_x ,
\end{equation}
where the populations are determined by the Wigner rotation matrix for a $\pi/2$ rotation:
\begin{equation}
|c_{m_x}|^2
=
\frac{1}{2^{2F}}
\binom{2F}{F+m_x}.
\label{eq:binomial_disc}
\end{equation}

For the experimentally relevant case $F=4$ (Cs),
\begin{equation}
\mathcal{P}(m_x)
=
\frac{1}{256}
\left\{
1,8,28,56,70,56,28,8,1
\right\},
\end{equation}
corresponding to
$m_x=-4,\dots,+4$.
Thus, when the state is maximally oriented along $y$, along the $x$ axis it possesses a strong quadrupole anisotropy concentrated near $m_x=0$, corresponding to negative alignment with respect to both of them.

\medskip

\textcolor{MyColor}{This conclusion can be generalized. 
Consider now an ensemble in the stretched state
\begin{equation}
|F,m_x=F\rangle ,
\end{equation}
where the quantization axis is chosen along the $x$ direction, in
accordance with the convention adopted throughout this work.
For this state,
\begin{align}
\langle F_x\rangle &=F,\\
\langle F_x^2\rangle &=F^2,\\
\langle F_y\rangle &=\langle F_z\rangle =0 .
\end{align}}

\textcolor{MyColor}{Using
\begin{equation}
F_x^2+F_y^2+F_z^2=F(F+1),
\end{equation}
and the symmetry between the transverse directions, one obtains
\begin{equation}
\langle F_y^2\rangle
=
\langle F_z^2\rangle
=
\frac{F}{2}.
\end{equation}}

\textcolor{MyColor}{The Cartesian alignment tensor is defined as
\begin{equation}
Q_{ij}
=
\frac12
\left\langle
F_iF_j+F_jF_i
\right\rangle
-
\frac13F(F+1)\delta_{ij}.
\end{equation}}

\textcolor{MyColor}{For the stretched state, the nonzero diagonal components are therefore
\begin{align}
Q_{xx}
&=
F^2-\frac13F(F+1)
=
\frac13F(2F-1),
\\
Q_{yy}
&=
Q_{zz}
=
\frac{F}{2}
-\frac13F(F+1)
=
-\frac16F(2F-1),
\end{align}
whereas all off-diagonal components vanish. This means that the component of the alignment tensor parallel to the orientation in the extended state is always positive, and the perpendicular ones are negative.}

\textcolor{MyColor}{Using the relation between Cartesian and irreducible tensor
representations,
\begin{equation}
\rho^{(2)}_0
=
\frac1{\sqrt6}
\left(
2Q_{xx}-Q_{yy}-Q_{zz}
\right),
\end{equation}
one obtains
\begin{equation}
\rho^{(2)}_0
=
\frac{F(2F-1)}{\sqrt6}.
\end{equation}
}

\textcolor{MyColor}{The relation between the orientation and the associated alignment has a
simple interpretation. A stretched state possesses cylindrical symmetry about its
axis $\mathbf n$. Consequently, its Cartesian alignment tensor can be written in the general form
\begin{equation}
Q_{ij}
=
Q_\perp\delta_{ij}
+
(Q_\parallel-Q_\perp)n_i n_j ,
\end{equation}}

\noindent\textcolor{MyColor}{where $Q_\parallel$ and $Q_\perp$ are the tensor components parallel and
perpendicular to the symmetry axis, respectively. For the stretched state oriented along $x$,
\begin{equation}
Q_\parallel=Q_{xx},\qquad
Q_\perp=Q_{yy}=Q_{zz},
\end{equation}
and therefore
\begin{equation}
Q_{xy}
=
(Q_{xx}-Q_{yy})n_x n_y .
\label{eq:Qxy_general}
\end{equation}}

\textcolor{MyColor}{For a small deviation of the orientation vector from the $x$ axis,
$n_x\simeq1$ and $n_y\simeq P_y$, where $P_y$ is the normalized
transverse orientation component. Hence,
\begin{equation}
Q_{xy}
\simeq
(Q_{xx}-Q_{yy})P_y .
\end{equation}}

\textcolor{MyColor}{Since
\begin{equation}
Q_{xx}-Q_{yy}
=
\frac12F(2F-1),
\end{equation}
the alignment component generated by a rigidly rotating spontaneous
polarization obeys
\begin{equation}
A_P\propto Q_{xy}\propto P_y ,
\end{equation}
where the proportionality coefficient depends on the normalization
convention adopted for the alignment variables in the main text.}

\textcolor{MyColor}{This geometrical relation also explains the symmetry of the
observed signal. The off-diagonal tensor component responsible for polarization rotation is proportional to the product of the two orientation projections,
\begin{equation}
Q_{xy}\propto n_x n_y .
\end{equation}
Consequently, the signal changes sign upon reversal of either transverse component of the orientation vector and vanishes when the orientation is exactly parallel to either the $x$ or $y$ axis. This symmetry corresponds to that observed for the alignment signal in our experiment.}

\bibliography{bibl}
\end{document}